\documentclass[a4paper]{article}

\usepackage[english]{babel}
\usepackage[utf8x]{inputenc}
\usepackage[T1]{fontenc}

\usepackage[numbers,sort&compress]{natbib}

\usepackage[a4paper,top=3cm,bottom=2cm,left=3cm,right=3cm,marginparwidth=1.75cm]{geometry}

\usepackage{amsmath}
\usepackage{graphicx}
\usepackage[colorinlistoftodos]{todonotes}
\usepackage[colorlinks=true, allcolors=blue]{hyperref}

\usepackage{authblk}

\title{ArCLight - a Compact Dielectric Large-Area Photon Detector}

\author{M.~Auger, Y.~Chen, A.~Ereditato, D.~Goeldi~,  I.~Kreslo\thanks{Corresponding author: igor.kreslo@cern.ch}, D.~Lorca, M.~Luethi, T.~Mettler, J.~R.~Sinclair, and M.~S.~Weber\\Laboratory for High Energy Physics\\Albert Einstein Center for Fundamental Physics\\Universit\"{a}t Bern, Switzerland}

\begin{document}
\maketitle

\begin{abstract}
ArCLight is a novel device for detecting scintillation light over large areas with Photon Detection Efficiency (PDE) of the order of a few percent. 
Its robust technological design allows for efficient use in large-volume particle detectors, such as Liquid Argon Time Projection Chambers (LArTPCs) or liquid scintillator detectors. 
Due to its dielectric structure it can be placed inside volumes with high electric field. 
It could potentially replace vacuum PhotoMultiplier Tubes (PMTs) in applications where high PDE is not required. The photon detection efficiency for a $10\times10$~cm$^2$ detector prototype was measured to be in the range of 0.8\% to 2.2\% across the active area.
\end{abstract}

\section{Introduction}
Modern detectors for neutrino oscillation physics are, to a large extent, represented by LArTPCs. 
At LHEP, University of Bern, an extensive LArTPC R\&D program is ongoing since 2007.
Our findings from this research program~\cite{ArT1,ArT2,HV} brought us to the ArgonCube project~\cite{ArC}. ArgonCube, a modular single-phase LAr TPC, is a leading contender for the LAr component of the DUNE near detector complex~\cite{DUNE}. The function of the ArgonCube Light readout system (ArCLight)
is to detect and quantify parameters of the prompt scintillation light, to provide t0 for the TPC, and give additional information on deposited energy.

One of the assets of the ArgonCube design is its extremely low fraction of insensitive argon volume (below 10\%). This applies strict limit to the volume that a light readout system can occupy, as well as its arrangement inside the detector. Classic vacuum PMT-based light readout systems require much larger volumes occupied by PMTs, located on low-potential side of the TPC. Such a solution is not compatible with the modular design of ArgonCube, where the TPC modules stand side-by-side and any additional space between them degrades tracking and calorimetric performance of the detector as a whole. 

One more specialty of LAr scintillation is its short wavelength: its emission spectrum is peaked at 128 nm. 
This spectrum is usually first shifted towards the visible range with a wavelength shifter (WLS), for which 1,1,4,4-tetraphenyl-1,3-butadiene (TPB) is a commonly used material. 
Strictly speaking, the process of converting 128 nm photons to visible ones (430 nm for TPB) is closer to scintillation than to fluorescence, because of the high energy of the initial photons. 
Absorption of these photons may lead to photo-ionization and dissociation rather than just excitation. 
This explains why the TPB conversion efficiency for LAr scintillation spectrum exceeds unity \cite{GEHMAN2011116}.

Several interesting low-volume large-area solutions have been investigated by the detector design community in order to meet the requirements of detecting  LAr scintillation light~\cite{Howard:2017,Moss:2014,Ignarra:2012,605551017,605551019,605551025,Whit16}.
However, these solutions are based on full reflection from the interface between a polymer and surrounding LAr, which is limited by the small difference in the refractive index. 
In the case of organic liquid scintillator as detector medium, this difference may practically vanish, rendering such solutions useless. 

The design of ArCLight was inspired by the ARAPUCA light trap sensor~\cite{Machado:2016jqe}. 
The ARAPUCA functions by trapping photons inside a cavity with highly reflective internal surfaces. 
After conversion to blue region with the first WLS, LAr scintillation light enters the cavity through a dichroic film. 
A second WLS coats the inner surface of the film. 
The latter shifts the wavelength beyond the transparent cutoff of the dichroic film, trapping the light within the cavity. 
ArCLight employs a similar principle, with the cavity void replaced by a solid transparent polymer sheet doped with a WLS dye. 
This improvement makes the detector substantially more robust, compact and stable, especially when scaling up to large areas.

\section{Detector Design}

\begin{figure}[ht!]
\centering
\includegraphics[width=0.45\textwidth]{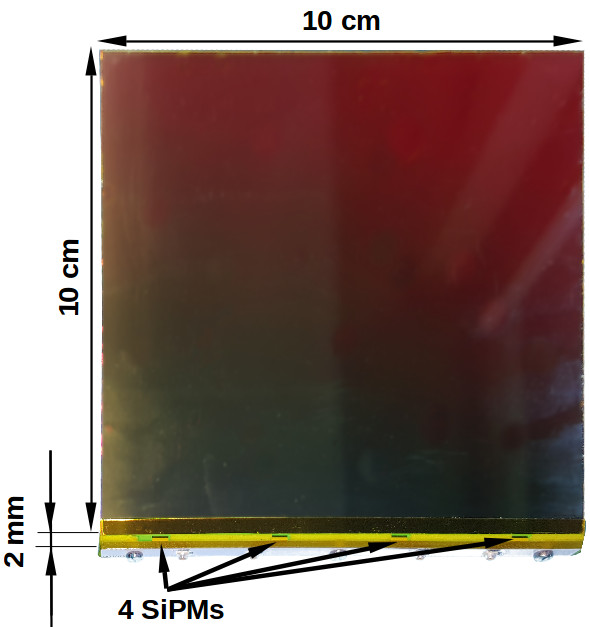}
\caption{ ArCLight $10\times10$~cm$^2$ prototype. Four silicon photomultipliers (SiPMs) in surface mount device (SMD) packages can be seen at the lower side, soldered to a narrow printed circuit board (PCB) which provides SMD coaxial connectors for signal readout. The rest of the sensor area is dielectric.}
\label{fig:acl}
\end{figure}

An ArCLight prototype with a 10$\times$10 cm$^2$ active area (98\% of total area) is shown in Figure~\ref{fig:acl}.

\begin{figure}[ht!]
\centering
\includegraphics[width=0.9\textwidth]{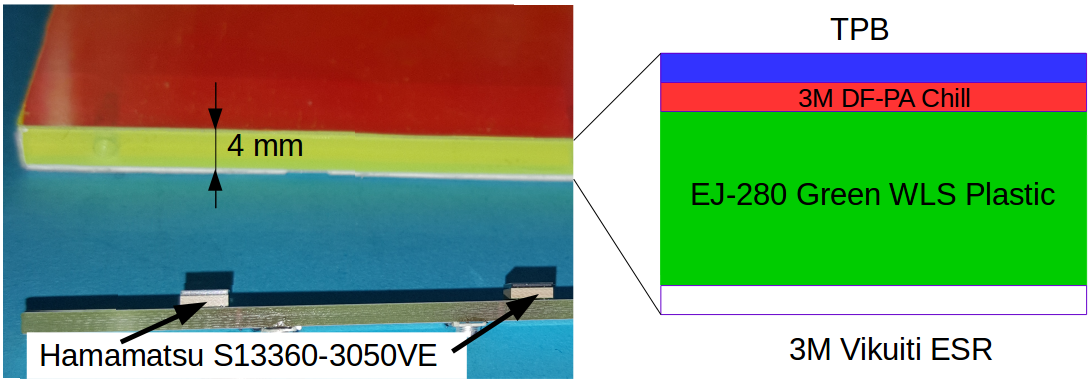}
\caption{The ArCLight light collector cross section near the corner. Two out of four SiPMs soldered to their carrier PCB are also shown, separate from the light collector.}
\label{fig:acl1}
\end{figure}

The structure of the device is shown in Figure~\ref{fig:acl1}. 
It is composed of a 4 mm thick EJ280 WLS\footnote{http://www.eljentechnology.com/products/wavelength-shifting-plastics/ej-280-ej-282-ej-284-ej-286} plate, with reflective films laminated to each side. 
The back face and edges are covered with a dielectric specular reflector foil\footnote{VM2000, former name for Vikuiti ESR, 3M Inc}, which has $\approx$98\% reflectance for the visible light spectrum~\cite{VIKUITI}. 
The front face is coated with a dichroic mirror\footnote{DF-PA Chill, 3M Inc.} film that is transparent in the blue and has high reflectance in the green range of the spectrum. 
Spectral characteristics of the dichroic film are measured for two incidence angles: 0$^{\circ}$ and 45$^{\circ}$ with a spectrophotometer\footnote{LAMBDA™ 750 UV/Vis/NIR spectrophotometer, PerkinElmer}.
Both films are held in place by thin layers of transparent adhesive. 
In a version aimed for use in LAr, an additional layer of TPB WLS is deposited on the front face to convert the scintillation light into the blue wavelength~\cite{Francini:2013lua}. 
At one edge of the WLS plate four Hamamatsu S13360-3050VE\footnote{http://www.hamamatsu.com/us/en/product/category/3100/4004/4113/S13360-3050PE/index.html} Silicon Photo-Multipliers (SiPMs) are mounted, each with a 3$\times$3 mm$^2$ sensitive area.

\begin{figure}[ht!]
\centering
\includegraphics[width=0.7\textwidth]{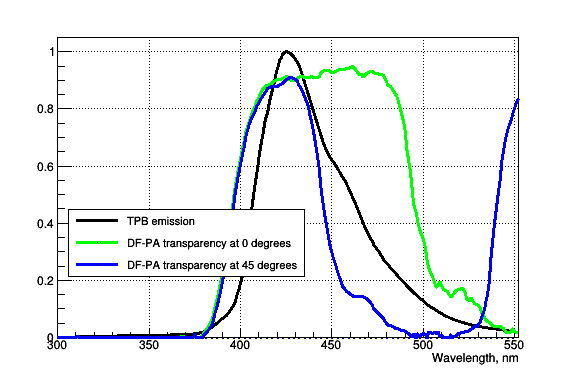}
\caption{TPB emission spectrum \cite{Francini:2013lua} (in a.u.) and the dichroic mirror transparency overlap for a substantial range of incidence angle. }
\label{fig:bluetop}
\end{figure}

\begin{figure}[ht!]
\centering
\includegraphics[width=0.7\textwidth]{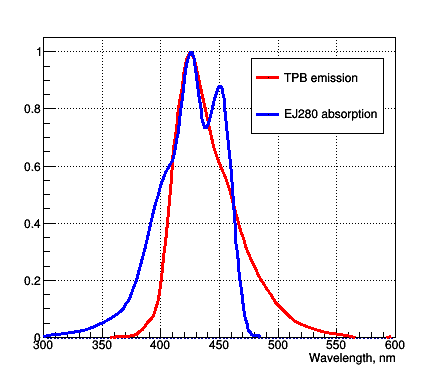}
\caption{TPB emission spectrum \cite{Francini:2013lua} and WLS absorption spectrum \cite{EJ280} (both in a.u.) are well matched.}
\label{fig:bluebot}
\end{figure}

\begin{figure}[ht!]
\centering
\includegraphics[width=0.7\textwidth]{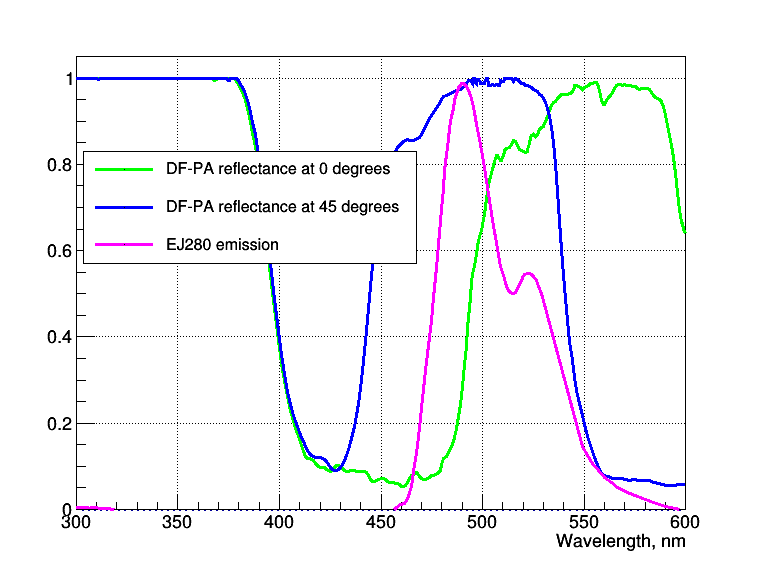}
\caption{WLS emission spectrum \cite{EJ280} (in a.u.) and the dichroic mirror reflectance fully overlap for angles above 10 degrees and partially below this value. About 30\% of the emitted light (below $\sim$10$^{\circ}$ incidence) is lost.}
\label{fig:greentop}
\end{figure}

\begin{figure}[ht!]
\centering
\includegraphics[width=0.7\textwidth]{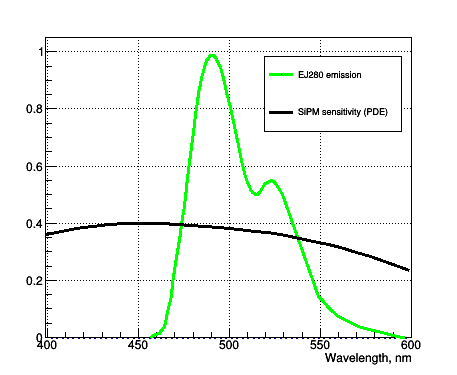}
\caption{ Spectral relationships:  WLS emission spectrum \cite{EJ280} (in a.u.) and SiPM spectral dependence of PDE \cite{MPPC}.}
\label{fig:greenbot}
\end{figure}

Photons in the spectral range of 400-450~nm pass through the dichroic reflector at the front face (Figure~\ref{fig:bluetop}) and are efficiently absorbed by WLS (Figure~\ref{fig:bluebot}). 
The latter re-emits them in green with a spectral maximum of 490 nm, as shown in Figure~\ref{fig:greentop} (WLS spectral parameters are taken from~\cite{EJ280}). 
These photons get trapped inside the WLS volume since the reflection coefficient of the dichroic film reaches 98\% for incidence angles above $\sim$10$^{\circ}$ at this wavelength, as shown in Figure~\ref{fig:greentop}. 
About 30\% of the emitted light (below $\sim$10$^{\circ}$ incidence) is therefore lost. 
However, due to specular reflectance of the mirror films, photons that are reflected once stay captured inside the WLS volume.
Eventually the photons reach openings in the side mirrors, where they are detected by the SiPMs. 
The emission spectrum of the WLS plastic matches the spectral sensitivity of the SiPMs~\cite{MPPC} closely, as shown in Figure~\ref{fig:greenbot}.

The expected PDE is calculated by folding the individual efficiencies at each step of photon transport:

\begin{equation}
\large  Q_{PDE} = \frac{1}{2}~\varepsilon_{TPB}~T_{425}~ \varepsilon_{WLS}~\varepsilon_{SA}~\varepsilon_{coll}~\varepsilon_{SiPM} 
\end{equation}

$\frac{1}{2} \varepsilon_{TPB}$ represents the efficiency of converting photons from 128~nm to 425~nm \cite{GEHMAN2011116}, taking into account the 2$\pi$ solid angle acceptance (to one side only). 
$T_{425}$ is the average transparency of the dichroic filter around 425~nm.
$\varepsilon_{WLS}$ is the conversion efficiency of the EJ280 WLS.
$\varepsilon_{SA}$ represents the average spectral acceptance of the reflecting surfaces; since the spectral cutoff for the dichroic film is angle-dependent, this acceptance is obtained by averaging the convolution of the WLS emission spectrum with the spectral reflectance of the dichroic film over the range of incidence angles. This value is estimated visually to a very rough precision, more accurate calculations can be done by folding all relevant spectral characteristics, but this is only viable if the angular distribution of the incoming photons is known.
$\varepsilon_{coll}$ is the efficiency of the reflecting surfaces to deliver photons to the SiPM openings. 
The value of $\varepsilon_{coll}$ can be estimated analytically~\cite{Segreto:2012} and is expressed by the surface reflection coefficient $R_{490}$ at a given wavelength (490~nm) and the fraction of surface covered by the SiPM active area $f=S_{det} / S_{tot}$, where $S_{tot}$ and  $S_{det}$ are total detector surface area and area covered by SiPMs, respectively:

\[\varepsilon_{coll} = \frac{f}{1-<R_{490}>(1-f)} \]

The main components of the final PDE are shown in Table 1. For these calculations the attenuation of green light inside the bulk EJ280 WLS material is neglected, since the attenuation length is of the order of meters~\cite{Howard:2017}. Unless the source is sited, values and their uncertainties are obtained by estimation from the measured spectral curves. Uncertainty of the $Q_{PDE}$ is obtained by error propagation.

\begin{figure}[htp!]
\centering
\includegraphics[width=0.9\textwidth]{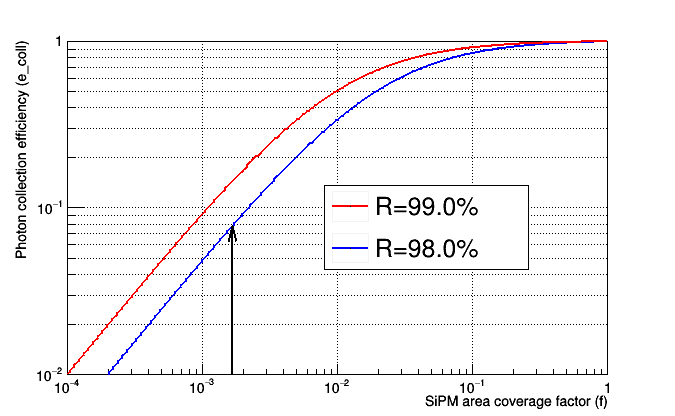}
\caption{Collection efficiency as a function of SiPM area coverage fraction f. The 98\% reflection coefficient is that used in the prototype \cite{VIKUITI}, 99\% shows a potential improvement.
The arrow indicates the value of the SiPM coverage for the prototype.}
\label{fig:colef}
\end{figure}

\begin{table}
\begin{center}
\begin{tabular}{ ||c|c||c|c|| } 
 \hline
 Parameter & Value & Parameter & Value\\ 
 \hline
 $\varepsilon_{TPB}$ & 1.2$\pm$0.1 \cite{GEHMAN2011116} & $T_{425}$ & 0.87$\pm$0.05\\ 
 \hline
 $\varepsilon_{WLS}$ & 0.86 \cite{EJ280} & $\varepsilon_{SA}$ &  0.7$\pm$0.4\\ 
 \hline
 $S_{det}$ & 0.36 cm$^2$ & $S_{tot}$ & 216 cm$^2$ \\ 
 \hline
 
 $\varepsilon_{coll}$ & 0.077 & $\varepsilon_{SiPM}$ & 0.38$\pm$0.02 \cite{MPPC}\\ 
 \hline
 $R_{490}$ & 0.98$\pm$0.01 & $Q_{PDE}$ & 0.007$\pm$0.004 \\ 
 \hline

\end{tabular}
  \\[10pt] 
  
\small{{\bf Table 1.} Summary of parameters and their values used in PDE estimation.} 

 \end{center}
\end{table}
The collection efficiency $\varepsilon_{coll}$ scales with the SiPM coverage area $f$, as shown in Figure~\ref{fig:colef}. The arrow indicates the value of the coverage for the prototype under study. Two curves are shown, one for a reflection coefficient of 98\%, as used in the studied prototype (see Figure \ref{fig:greentop} and \cite{VIKUITI}), and one for a potentially improved setup with 99\% (better spectral match of dichroic film to the WLS emission).

\section{Measurements}

Making a stable source of light pulses containing $\sim$100 photons is a challenging task.
For this purpose a combination of a $^{241}$Am radioactive source and a thin scintillating foil was used. 
The source emits $\alpha$-particles with mean energy of $\approx$3~MeV (the active material is embedded inside a Rhodium matrix). 
The 0.5~mm thick NE102A (EJ212) scintillating foil ($\lambda_{max}$=423~nm \cite{EJ212}) was brought into contact with the source surface.
The intensity of light pulses were calibrated with a Philips 56AVP vacuum PMT, with a known quantum efficiency of $QE$=22$\pm2\%$ at 423~nm. 

The peak voltage of the PMT output pulse is proportional to number of photoelectrons, $N_{pe}$, in a pulse, assuming pulse shape is constant: $V=G N_{pe}$. 
The single photo-electron voltage response, $G$, was calibrated with a pulsed LED using the technique described in~\cite{PMT}. 
The ideal Poisson variance of the PMT output voltage signal is defined by $\sigma_V^2=G^2 N_{pe}$. 
In reality, this value is approximately four times larger for a typical PMT~\cite{PMT}, due to contributions from an excess noise factor and several other effects. 
Therefore, $G$ was calculated as $G=\frac{1}{4}\sigma_V^2/V$ for several voltages and the mean taken. 
The obtained value of $G$=3.3$\pm$0.3~mV/p.e was validated by single-photo-electron response method, observing signal from very low light pulse intensity (<1 p.e.).

\begin{figure}[h!]
\centering
\includegraphics[width=0.5\textwidth]{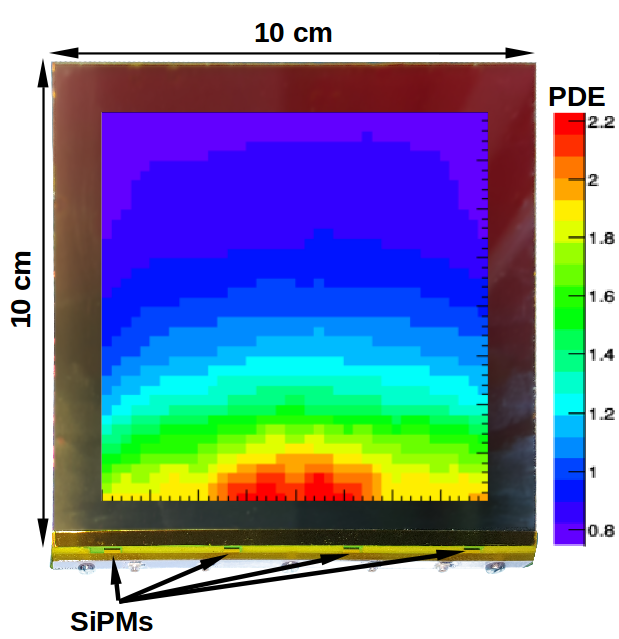}
\caption{ ArCLight measured photon detection efficiency (PDE) map overlaid onto the $10\times10$~cm$^2$ surface of the detector. Color scale is shown in \% units. Increase in PDE near SiPM side is due to increased fraction of direct photons, avoiding initial loss at first reflection from the dichroic mirror. Short component of WLS bulk attenuation is likely contributing to this direct access as well.}
\label{fig:10x10}
\end{figure}

Using this PMT, the intensity of the light pulses from the scintillator based light source was estimated to be at the level of 700$\pm$100 photons per pulse. 
This number was used to derive the PDE of the $10\times10$~cm$^2$ prototype detector. 
The results of this measurement are shown in Figure~\ref{fig:10x10}, where the PDE distribution over the $10\times10$~cm$^2$ area of the detector is plotted.
The measurements were conducted at room temperature. At 87K, in LAr, a small drift in spectral characteristics of dichroic film is expected due to mechanical shrinkage of the reflecting stack structure. However, this shift is limited to about 2\% (assuming thermal expansion factor of less than $10^{-4}K^{-1}$). Such a change should not lead to a noticeable variation of the PDE, withing given measurement uncertainty.


\section{Conclusions}
A novel design for a large-area photon detector is proposed. The ArCLight device is made of dielectric materials, except a single edge where SiPMs are mounted. The thin and robust structure with approaching 100\% active area allows efficient integration into large LArTPCs, as well as other liquid scintillator based detectors. The photon detection efficiency for a $10\times10$~cm$^2$ detector prototype was measured to be of the order of 1\% (in the range of 0.8\% to 2.2\% ), which has shown good agreement with theoretical expectations. The design is expected to be easily scalable to several square meters of surface area due to long (order of meters) attenuation length of the used WLS plastic. The prototype was found to be efficient in the wavelength range of 400-460~nm. However, the design could be adjusted for different spectral ranges by a suitable choice of materials. 


\bibliographystyle{unsrtnat}
\bibliography{sample.tex}

\end{document}